\documentclass[pra, preprint, superscriptaddress]{revtex4}
\usepackage{graphicx}
\usepackage{mathrsfs}
\usepackage{dcolumn}
\usepackage{amsmath}
\usepackage{amssymb}
\usepackage{color}
\usepackage{float}
\usepackage{epstopdf}

\def\beq{\begin{equation}}
\def\eeq{\end{equation}}

\def\simg{\,\hbox{\kern.1em \lower.6ex \hbox{$\sim$} \kern-1.12em
          \raise.6ex \hbox{$>$} }}
\def\siml{\,\hbox{\kern.1em \lower.6ex \hbox{$\sim$} \kern-1.12em
          \raise.6ex \hbox{$<$} }}
\def\Xint#1{\mathchoice
   {\XXint\displaystyle\textstyle{#1}}%
   {\XXint\textstyle\scriptstyle{#1}}%
   {\XXint\scriptstyle\scriptscriptstyle{#1}}%
   {\XXint\scriptscriptstyle\scriptscriptstyle{#1}}%
   \!\int}
\def\XXint#1#2#3{{\setbox0=\hbox{$#1{#2#3}{\int}$}
     \vcenter{\hbox{$#2#3$}}\kern-.5\wd0}}

\def\dashint{\Xint-}
\newcommand\bea{\begin{eqnarray}}
\newcommand\eea{\end{eqnarray}}

\begin{document}

\title{On the asymptotic prime partitions of integers}
\author{Johann Bartel}
\affiliation{Institut Pluridisciplinaire Hubert Curien, Physique Th\'eorique,
Universit\'e de Strasbourg, F-67037 Strasbourg, France}
\email{Johann.Bartel@iphc.cnrs.fr}
\author{R. K. Bhaduri}
\affiliation{Department of Physics and Astronomy, McMaster University,
Hamilton L8S4M1, Canada}
\email{bhaduri@physics.mcmaster.ca}
\author{Matthias Brack}
\affiliation{Institute of Theoretical Physics, University of Regensburg,
D-93040 Regensburg, Germany}
\email{matthias.brack@ur.de}
\author{M. V. N. Murthy}
\affiliation{The Institute of Mathematical Sciences, Chennai 600 113, India}
\email{murthy@imsc.res.in}

\begin{abstract}

In this paper we discuss ${\cal P}(n)$, the number of ways a given integer $n$ may be 
written as a sum of primes. In particular, an asymptotic form ${\cal P}_{as}(n)$ valid for 
$n\rightarrow\infty$ is obtained analytically using standard techniques of quantum 
statistical mechanics. First, the bosonic partition function of primes, or the generating 
function of unrestricted prime partitions in number theory, is constructed. Next, the 
density of states is obtained using the saddle-point method for Laplace inversion of the 
partition function in the limit of large $n$. This directly gives the asymptotic number of 
prime partitions ${\cal P}_{as}(n)$. The leading term in the asymptotic expression grows 
exponentially as $\sqrt{n/\!\ln(n)}$ and agrees with previous estimates. We calculate the 
next-to-leading order term in the exponent, porportional to $\ln[\ln(n)]/\!\ln(n)$, and show 
that an earlier result in the literature for its coefficient is incorrect. Furthermore, we 
also calculate the next higher order correction, proportional to $1/\!\ln(n)$ and given in 
Eq.\ (\ref{partas}), which so far has not been available in the literature. Finally, we 
compare our analytical results with the exact numerical values of ${\cal P}(n)$ up to $n 
\sim 8\cdot 10^6$. For the highest values, the remaining error between the exact ${\cal P}(n)$ 
and our ${\cal P}_{as}(n)$ is only about half of that obtained with the leading-order (LO) 
approximation. But we also show that, unlike for other types of partitions, the asymptotic 
limit for the prime partitions is still quite far from being reached even for $n \sim 10^7$. 

\bigskip

\end{abstract}

\maketitle

\section{Introduction}

Consider $N$ identical ideal bosons that occupy a quantum system with 
equispaced single-particle energy levels at integer-valued $n$, with the 
lowest level at $n=0$. This is simply the one-dimensional harmonic oscillator 
spectrum with the zero-point energy set to zero. In the ground state of this 
system, all the bosons occupy the lowest level at $n=0$. When a large excitation
energy is given to the system, there are many ways by which 
this energy can be distributed amongst the $N$ bosons. In fact, this 
is precisely the number ${\cal P}(n)$ of ways in which an integer $n$ can 
be partitioned into a sum of integers less than or equal to $n$. The asymptotic 
form of ${\cal P}(n)$ (corresponding to $N\rightarrow \infty$ particles) is 
precisely the Hardy-Ramanujan formula \cite{HR} for the number partitions. The 
generating function in number theory is intimately connected to the bosonic partition 
function of statistical mechanics. It is interesting to note that this was written 
down by Hardy and Ramanujan years before the Bose-Einstein distribution was 
discovered in physics. In an earlier publication by some of the present authors 
\cite{muoi}, the asymptotic quantum density of states $\rho(E)$ was shown to be 
the ${\cal P}(n$=$E)$ known from number theory. This was done by performing the inverse 
Laplace transformation of the partition function using the saddle-point method.

It is obvious that the same technique of statistical mechanics may be applied to obtain any 
partition of a positive integer $n$, thus in particular also for its partition into {\it primes} 
$p$, if we start with a system whose single-particle levels are simply the primes $p$. 
The total energy now is given as a sum of primes, and the corresponding density of 
states is given by the {\it number of prime partitions} ${\cal P}(n)$. For our calculations, 
we require to convert the sum over primes into a continuous integral. For this, we need the 
density of primes which may be deduced from the well-known prime number theorem (see the
Appendix). The leading-order (LO) analytical expression for ${\cal P}(n)$ in the asymptotic limit 
$n \rightarrow \infty$ is available in the literature \cite{roth,yang}. Corrections to the LO 
asymptotic result have been derived by Vaughan \cite{vaughan} using the saddle-point method. 
While our LO result, multiplied by a pre-exponential factor, agrees with that given by Vaughan 
\cite{vaughan}, our next-to-leading order (NLO) term in the exponent has a different coefficient 
($-\frac12$) from that given by Vaughan (+1) which we are convinced is incorrect. 
Furthermore, while only an error estimate was given in Ref.\ \cite{vaughan} for
the remaining terms beyond the NLO correction, we give a precise analytical expression 
for the next higher-order correction in the exponent of ${\cal P}(n)$. Our asymptotic 
result, which is denoted by ${\cal P}_{as}(n)$ in Eq.\ (\ref{partas}), is compared 
numerically with the exactly computed ${\cal P}(n)$. Although our asymptotic form comes 
much closer to the true ${\cal P}(n)$ than that of Ref.\ \cite{vaughan} for large $n$, 
we find that all asymptotic expressions discussed here are still quite far from reaching 
the exact ${\cal P}(n)$, even for as large numbers as $n \sim 10^7$. The reason for this 
slow approach to the asymptotic form will be discussed after presenting the numerical results.

The plan of our paper is as follows. In Section II.A, we present our tools of statistical 
mechanics for a system whose single-particle spectrum is given by distinct primes $p$ 
and whose total energy $E$ is distributed amongst $N$ bosonic particles. In Section 
II.B, an analytical asymptotic form of the canonical bosonic partition function $Z(\beta)$ 
is derived and checked by an exact numerical computation of $Z(\beta)$. In Section II.C, 
we obtain the many-body density of states $\rho(E)$ by Laplace inversion of $Z(\beta)$ using 
the saddle-point approximation. The next two subsections II.D and II.E are devoted to 
deriving our main result ${\cal P}_{as}(n)$ for the numbr of prime partitions, as given in Eq.\ 
(\ref{partas}), whereby the continuous energy variable $E$ is identified with the discrete 
number $n$. In Section III, our asymptotic result, as well as other expressions, are compared 
numerically with the exact function ${\cal P}(n)$ for the prime partitions which we have 
computed up to $n \sim 8\cdot 10^6$. We conclude our paper with a short summary in Section IV. 
Some details about the density of primes, relevant to our analysis, are presented in the Appendix.

\section{Partitions into primes}

\subsection{$N$-body system with a single-particle spectrum of primes}

We set out to formulate our method for a fictitious bound system whose discrete, non-degenerate 
single-particle energies are given by the primes $p=2, 3, 5, \dots$ We do not know of any physical 
system having this property (see also the last paragraph of our conclusions). But the use of 
quantum-statistical methods together with semiclassical ``trace formulae'' \cite{gutz,book} to 
purely mathematical spectra can be very enlightening. A famous example is the spectrum of the 
non-trivial zeros of the Riemann zeta function. The quest for a Hamiltonian with this spectrum 
(see Ref.\ \cite{bend} for a recent attempt) has motivated the research of many physicists and 
mathematicians, and may even be hoped to lead to a proof of the Riemann hypothesis. (For two nice 
reviews about this topic, see Refs. \cite{berry,bohigas}.) A trace formula for the prime spectrum 
is given in the Appendix.

Consider now a large number $N$ of bosonic particles occupying these levels described by the prime
spectrum. The total energy $E$ of the system is given by 
\beq 
E=\sum_{p} n_p\, p\,.
\label{Etot}
\eeq 
Here and in the following, the sums $\sum_p$ run over all primes, and $n_p$ are the occupancies 
of the levels which may be zero or positive integers such that
\beq
\sum_p n_p = N\,.
\label{Nsum}
\eeq 
In other words, the total energy $E$ in (\ref{Etot}) is given by any of its partitions into primes,
restricted by the particle number conservation (\ref{Nsum}). The number of possible such partitions 
shall be denoted by ${\cal P}_N(E)$, where the subscript $N$ keeps track of the number of particles. 
Although $E$ is therefore necessarily integer, we treat it as a continuous variable like in statistical 
mechanics. It is important to realize that ${\cal P}_N(E)$ is identical to the many-body density of 
states $\rho_N(E)$ that is related to the canonical $N$-body partition function $Z_N(\beta)$ by
\beq
Z_N(\beta) = \int_0^\infty dE\, \rho_N(E)\,\exp(-\beta E)\,,
\label{zcan}
\eeq
where $\beta=1/kT$ is the inverse temperature. Note that this expression, with 
$\rho_N(E)={\cal P}_N(E)$, has the familiar form of the generating function of partitions 
used in number theory \cite{HR}. Since Eq.\ (\ref{zcan}) formally is a Laplace transform,
the density of states $\rho_N(E)$ can be obtained from the partition function by its inverse
Laplace transform 
\beq
\rho_N(E) = \frac{1}{2\pi i}\int_{-i\infty}^{i\infty} d\beta\, Z_N(\beta)\,\exp(\beta E)\,.
\label{rhodef}
\eeq
We shall perform this Laplace inversion in the saddle-point approximation.

In terms of the single-particle spectrum, the canonical partition function $Z_N(\beta)$ may be written, 
after taking the limit $N\rightarrow \infty$, as
\beq
Z_\infty(\beta)=\prod_p \frac{1}{1-e^{-\beta\,p}}\,,
\label{zinfty}
\eeq
where the product runs over all primes $p$. For simplicity, we shall henceforth omit the subscript 
``$\infty$'' from $Z(\beta)$ as well as from the functions $\rho(E)$ and ${\cal P}(E)$. Having taken 
the limit $N\rightarrow \infty$ implies that the partitions of the total energy now are unrestricted, 
admitting any number of summands allowed by the value of the energy (\ref{Etot}). Taking the Laplace 
inverse of $Z(\beta)$ according to (\ref{rhodef}) thus leads to $\rho(E)={\cal P}(n$=$E)$ in the limit 
$N\to\infty$.

In doing the transform Eq.\ (\ref{rhodef}), we define the function
\beq
S(\beta)=\beta E + \ln Z(\beta)\,,
\label{sdef}
\eeq
which formally defines the canonical entropy. We now evaluate the inverse Laplace transform in Eq.\ 
(\ref{rhodef}) using the method of steepest descent, or saddle-point method. Hereby one is looking for
a stationary point $\beta_0$ of the function $S(\beta)$ appearing in the exponent of the inverse 
Laplace integral, which corresponds to a saddle point in the complex $\beta$ plane. This leads to the
{\it saddle-point equation} (or saddle-point condition)
\beq
\left. \frac{\partial S(\beta)}{\partial \beta}\right|_{\beta_0}
                   = E+\frac{Z'(\beta_0)}{Z(\beta_0)}=0\,.
\label{spc}
\eeq
If this equation has a solution $\beta_0$, which will be a function $\beta_0(E)$
of the energy, one evaluates the derivatives of $S(\beta)$ at $\beta_0$:
\beq
S^{(n)}(\beta_0)=\left. \frac{\partial^n S(\beta)}{\partial \beta^n}\right|_{\beta_0}.
\eeq 
The result of the inverse Laplace transform then is given by
\beq
\rho(E)=\frac{e^{S(\beta_0)}}{\sqrt{2\pi S^{(2)}(\beta_0)}}
\left[1+\cdots\right],
\label{rhosol}
\eeq
where the dots indicate so-called cumulants involving higher derivatives of the entropy, which become 
more important for large $\beta$ (see, e.g., Ref.\ \cite{jelovic}). Since we are interested here in the 
limit $\beta\to 0$ relevant for the asymptotics of large $E$, we shall neglect them. 


\subsection{Asymptotic partition function}

Taking the logarithm of the partition function (\ref{zinfty}) gives a sum over all
primes $p$, which we may also write as an integral 
\beq
\ln Z(\beta) = -\sum_p \ln (1-e^{-\beta p}) = -\int_{x_0}^{\infty} dx\,g(x) \ln(1-e^{-\beta x})\,,
\label{zex}
\eeq
where $g(x)$ is the exact density of primes given by the sum of delta function distributions
\beq
g(x) = \sum_p \delta(x-p)\,,
\label{gofx}
\eeq
and $x_0$ is any real number smaller than the lowest prime: $x_0<2\,$. The sum of distributions (\ref{gofx})
may be decomposed, as in many exact trace formulae for spectral distributions \cite{book}, into an 
average part $g_{av}(x)$, which is a continuous function of $x$, and an oscillating part representing 
itself as a sum of harmonics whose superposition results in the discreteness of the spectrum described 
by Eq.\ (\ref{gofx}). (See the Appendix for the case of the prime spectrum). The main object of the 
present study is the asymptotic behaviour of ${\cal P}(n)$ for large $n$, for which the use of $g_{av}(x)$ 
in (\ref{zex}) is sufficient. For large $x$, the discreteness of $g(x)$ may be ignored, and the 
resulting ${\cal P}_{as}(n)$ is a smooth function of $n$ as a continuous variable. 

We therefore now replace the exact $g(x)$ in (\ref{zex}) by the average prime density $g_{av}(x)$. In doing 
so, we define the logarithm of the average partition function
\beq
\ln Z_{av}(\beta) = -\int_{a}^{\infty} dx\,g_{av}(x) \ln(1-e^{-\beta x})\,,
\label{zav}
\eeq
where the constant $a$ must be chosen carefully as discussed in the following. As a specific choice of 
$g_{av}(x)$, we use the asymptotic prime density that is well-known from number theory (see the Appendix):
\beq 
g_{av}(x) = 1/\!\ln(x)\,.
\label{gavofx}
\eeq 
Since we are only interested in asymptotic results, it will be sufficient to look at the limit 
$\beta\rightarrow 0$, i.e., the high-temperature limit of the partition function. 

The integrand (\ref{gavofx}) in (\ref{zav}) has a pole at $x=1$, which becomes relevant when $a<1$. 
We therefore define the following principal-value integral
\beq
I(a,\beta) = - \lim_{\epsilon\to 0} \left[ \int_a^{1-\epsilon} dx\, \frac{1}{\ln(x)}\ln(1-e^{-\beta x})
                               + \int_{1+\epsilon}^\infty dx\, \frac{1}{\ln(x)}\ln(1-e^{-\beta x}) \right], 
                               \quad (a \neq 1)  
\label{Iab}
\eeq
which in the following is denoted by the symbol $\dashint_a^\infty dx (\dots)$, so that
\beq
\ln Z_{av}(a,\beta) = I(a,\beta)= - \dashint_a^\infty dx\, \frac{1}{\ln(x)}\ln(1-e^{-\beta x})\,.
\label{ZI}
\eeq
This integral exists for any $a\neq 1$ and for finite $\beta$. We now make the change of variable 
$y=\beta x$ to obtain
\beq
I(a,\beta) = \frac{1}{\beta\ln(\beta)}\,\dashint_{a\beta}^{\infty} dy\,
\frac{1}{\left[1-\frac{\ln(y)}{\ln(\beta)}\right]}\ln\left(1-e^{-y }\right).
\label{Iab1}
\eeq
In order to make the next step more clear, we define
\beq
\tau = 1/\beta\,
\label{taub}
\eeq
and rewrite (\ref{Iab1}) as
\beq
I(a,\tau) = - \frac{\tau}{\ln(\tau)}\,\dashint_{a/\tau}^{\infty} dy\,
\frac{1}{\left[1+\frac{\ln(y)}{\ln(\tau)}\right]}\ln\left(1-e^{-y }\right),
\label{Iabtau}
\eeq
which we want to evaluate asymptotically in the high-temperature limit $\tau\to\infty$. 
We split it into two parts, writing
\beq
I(a,\tau) = - \frac{\tau}{\ln(\tau)}\left[\,
              \dashint_{a/\tau}^{\tau} dy\,
              \frac{1}{\left[1+\frac{\ln(y)}{\ln(\tau)}\right]}\ln(1-e^{-y})
            + \int_{\tau}^{\infty} dy\,
              \frac{1}{\left[1+\frac{\ln(y)}{\ln(\tau)}\right]}\ln(1-e^{-y})
              \,\right].
\eeq
If we fix $a$ to an arbitrary value in the limits $1 < a < 2$ and take $\tau >1$, we may 
approximate the first
integral by the first term of the binomial expansion of its denominator and write
\beq
I(a,\tau) \simeq - \frac{\tau}{\ln(\tau)}\!\left[
                 \int_{a/\tau}^{\tau}\! dy\,
                 \left[1-\frac{\ln(y)}{\ln(\tau)}\right]\ln(1-e^{-y})
               + \int_{\tau}^{\infty}\! dy\,
                 \frac{1}{\left[1+\frac{\ln(y)}{\ln(\tau)}\right]}\ln(1-e^{-y})
                 \right]\!.
\eeq
In the limit $\tau\to\infty$, the second integral goes to zero and the first integral
gives the asymptotic approximation
\beq
I_{as}(\tau) = - \frac{\tau}{\ln(\tau)}\int_{0}^{\infty} dy\,
                       \left(1-\frac{\ln(y)}{\ln(\tau)}\right)\ln(1-e^{-y})\,,
\label{Ias}
\eeq
which no longer depends on the value of $a$.
Using (\ref{taub}) and (\ref{ZI}), we obtain the following asymptotic form for the 
logarithm of the partition function, which we call $\ln Z_{as}(\beta)$ and which we 
can evaluate analytically:
\beq
\ln Z_{as}(\beta) = \frac{1}{\beta\ln(\beta)}\int_{0}^{\infty} dy\,
                    \left(1+\frac{\ln(y)}{\ln(\beta)}\right)\ln(1-e^{-y})
                  = -\frac{f_1}{\beta\ln(\beta)}+\frac{f_2}{\beta\ln^2(\beta)}\,, 
\label{Zas}
\eeq
where
\beq
f_1 = \frac{\pi^2}{6}, \qquad f_2 = \frac{C\pi^2}{6}+\sum_k\frac{\ln(k)}{k^2} = 1.88703\,. 
\eeq
Here $C=0.577216$ is the Euler constant and the sum over $k$ has been evaluated numerically
(with $k_{max}\sim 10,000$). 

\begin{figure}[h]
\centering
\vspace*{-1.25cm}
\hspace*{-0.6cm}\includegraphics[width=1.0\columnwidth,clip=true,angle=0]{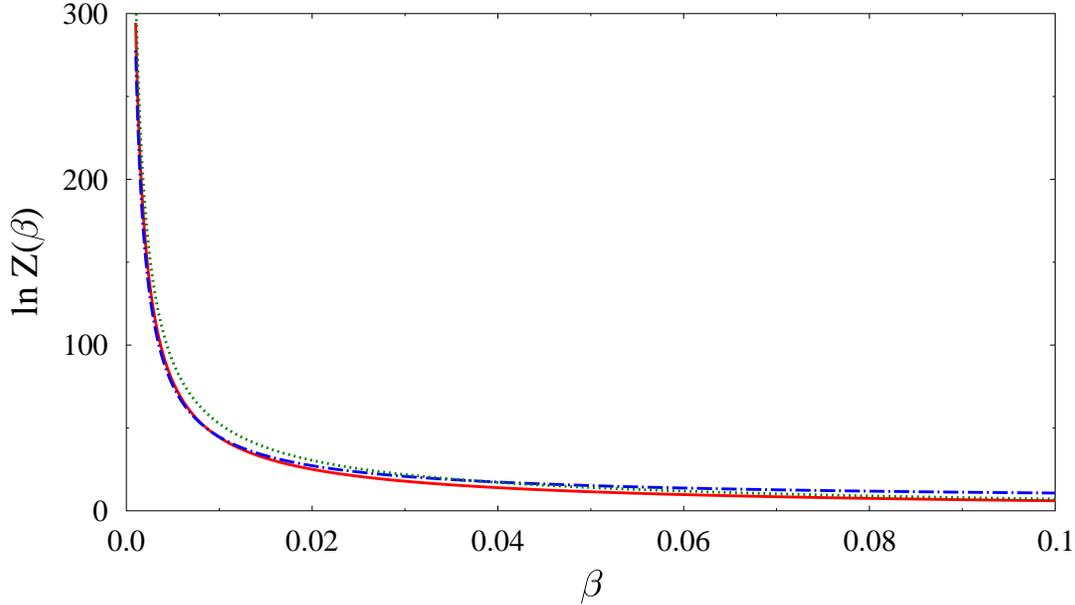}
\vspace*{-0.7cm}
\caption{Logarithm $\ln Z(\beta)$ of the partition function plotted versus $\beta$.
Solid line (red): exact function (\ref{zex}). Dotted line (green): numerically integrated
principal-value integral $\ln Z_{av}(a,\beta)$ in (\ref{ZI}) with $a=0$. Dash-dotted (blue) 
line: asymptotic approximation $\ln Z_{as}(\beta)$ in (\ref{Zas}).}
\label{figzofb1} 
\end{figure}

We now want to test the quality of the approximation (\ref{Zas}), which should become accurate
in the limit $\beta\to 0$. To that purpose we first integrate the principal-value integral 
$\ln Z_{av}(a,\beta)$ in (\ref{ZI}). Here we choose $a=0$ for definiteness; we emphasize that this 
choice is {\it a priori} independent of the fact that $1<a<2$ was used to derive the approximation 
(\ref{Zas}). Then we compare it to the exact function (\ref{zex}) and to the approximation 
$\ln Z_{as}(\beta)$ in (\ref{Zas}). The results are shown in Fig.\ \ref{figzofb1}. 

\begin{figure}[h]
\centering
\vspace*{-1.25cm}
\hspace*{-0.6cm}\includegraphics[width=1.0\columnwidth,clip=true,angle=0]{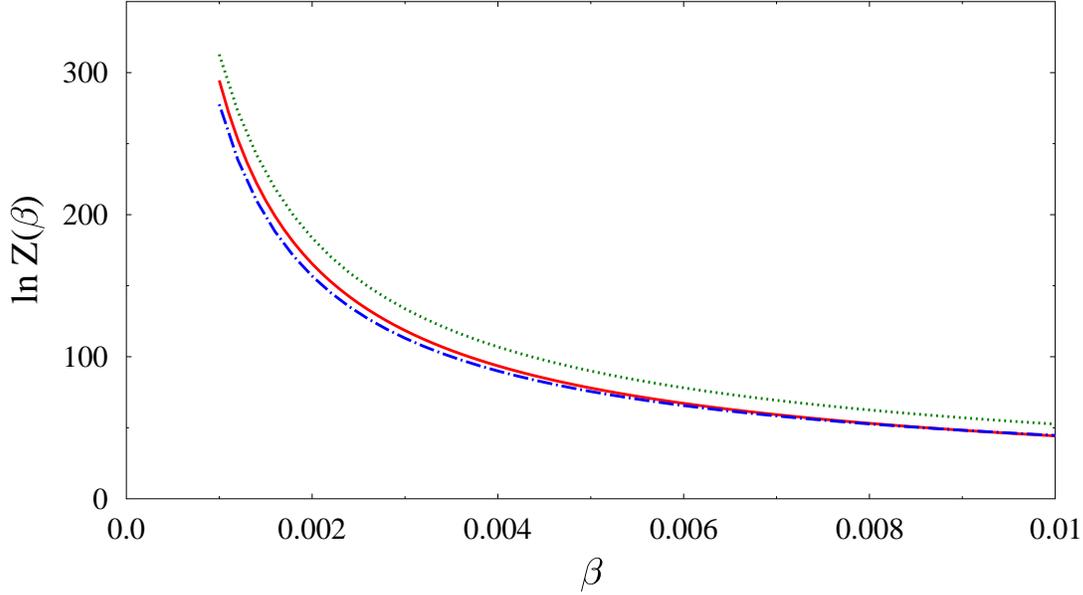}
\vspace*{-0.7cm}
\caption{Same as Fig.\ \ref{figzofb1}, but shown in limit of small $\beta$.}
\label{figzofb} 
\end{figure}

We see that both approximations 
approach the exact values closely for small $\beta$, while $\ln Z_{av}(\beta)$ is better than 
$\ln Z_{as}(\beta)$ for the largest values of $\beta$. In Fig.\ \ref{figzofb}, we see the same 
in a region of smaller values for $\beta$. The approximation $\ln Z_{as}(\beta)$ given in 
(\ref{Zas}) crosses the exact curve near $\beta \sim 0.008$ and appears to stay below it for 
$\beta\to 0$. It reveals itself as an excellent asymptotic approximation to the exact 
$\ln Z(\beta)$ in the small-$\beta$ limit.

Using the analytical form (\ref{Zas}) of the partition function, the inverse Laplace transform 
(\ref{rhodef}) can now be done in the saddle-point approximation as outlined at the end of Sect. II.A.



\subsection{Saddle-point approximation}

In order to find the saddle point $\beta_0$, we isolate the most singular terms in $S(\beta)$ in the 
high-temperature limit. We first write the entropy, using (\ref{Zas}) above, in the form
\beq
S(\beta)= \beta E-\frac{f_1}{\beta\ln(\beta)}
+\frac{f_2}{\beta\ln^2(\beta)}\,.
\label{slo}
\eeq
Since the entropy above is given up to order $1/\!\ln^2(\beta)$, all further calculations will be 
done up to this order. To begin with we need the following derivatives of the entropy
\beq
S^{(1)}(\beta)= E+\frac{f_1}{\beta^2\ln(\beta)} +\frac{f_1}{\beta^2\ln^2(\beta)} 
-\frac{f_2}{\beta^2\ln^2(\beta)}+\cdots,
\label{slo1}
\eeq
\beq
S^{(2)}(\beta)= -\frac{2f_1}{\beta^3\ln(\beta)}-\frac{3f_1}{\beta^3\ln^2(\beta)}
+\frac{2f_2}{\beta^3\ln^2(\beta)}+\cdots. 
\label{slo2}
\eeq
The saddle-point equation (\ref{spc}) can therefore be written in the following form
\beq
\beta E=-\frac{f_1}{\beta\ln(\beta)}
+\frac{f_2}{\beta\ln^2(\beta)}
-\frac{f_1}{\beta\ln^2(\beta)}
+\cdots. 
\label{spcond}
\eeq
This is a transcendental equation whose solution $\beta_0$ can be only obtained iteratively as outlined
in Sec.\ III.D below. However, we may use the above equation directly to write $S(\beta)$ in terms of 
the as yet undetermined $\beta_0$ as
\beq
S(\beta_0)= 2\beta_0 E+\frac{f_1}{\beta_0\ln^2(\beta_0)}+\cdots,
\eeq
and similarly, 
\beq
S^{(2)}(\beta_0)= \frac{1}{\beta_0^2}\left[2\beta_0 E
-\frac{f_1}{\beta_0\ln^2(\beta_0)}+\cdots\right]. 
\eeq
With Eq.\ (\ref{rhosol}) we obtain the asymptotic density of states as
\begin{equation}
\rho(E) = 
\frac{\exp(2\beta_0 E+\frac{f_1}{\beta_0\ln^2(\beta_0)}+\cdots)}
{\sqrt{(2\pi/\beta_0^2)\left[2\beta_0 E
-\frac{f_1}{\beta_0\ln^2(\beta_0)}+\cdots\right]}}.
\label{rhoprime}
\eeq
This is the same as the asymptotic expression for the number of prime partitions ${\cal P}(n)=\rho(E$=$n)$. 
Any further analysis requires the explicit solution $\beta_0(E)$, which we derive next.

\subsection{Saddle-point solution to leading order}
 
To leading order ${\cal O}[1/\ln(\beta_0)]$, the saddle-point equation (\ref{spcond}) reads 
\beq
\beta E=-\frac{f_1}{\beta\ln\beta}\,.
\label{spcondlo}
\eeq
We now solve this equation iteratively. Let $\tau=1/\beta$:
\beq
\frac{f_1}{E}=\frac{\ln(\tau)}{\tau^2}. 
\eeq
We start by assuming the solution to be of the form
\beq 
\tau=a_1E^{a_2}[\ln(E)]^{a_3}, 
\eeq
where $a_1$, $a_2$, and $a_3$ are constants to be determined using
Eq. (\ref{spcondlo}). Upon substitution, assuming large $E$, we get
\beq
\frac{f_1}{E}=\frac{1}{a_1^2 E^{2a_2}(\ln E)^{2a_3}}[\ln a_1 +a_2
  \ln(E) +a_3\ln\ln(E)]\approx \frac{a_2}{a_1^2 E^{2a_2}(\ln E)^{2a_3-1}}.
\label{trans}
\eeq
First we determine the leading term, comparing powers, to find the solutions
\beq
a_3 = \frac{1}{2}\,, \quad
a_2 = \frac{1}{2}\,, \quad
a_1^2 = \frac{a_2}{f_1}=\frac{3}{\pi^2}\,. 
\eeq
Thus we have the leading solution given by
\beq
\tau=\frac{1}{\beta_0}=
\sqrt{\frac{3}{\pi^2} E\ln(E)}\,.
\label{solnlo}
\eeq

To leading order, therefore, we have the following result for the density, 
or equivalently for unrestricted prime partitions:
\beq
\rho(E)=
\frac{e^{S(\beta_0) }}{\sqrt{2\pi S''(\beta_0)}}
=\frac{e^{2\pi\sqrt{E/[3\ln(E)]}}}{\sqrt{4E^{3/2}[3\ln(E)]^{1/2}}}\,.
\label{partlo}
\eeq
Apart from the prefactor, it is well known \cite{roth, yang} that 
$\ln[\rho(E)]\approx 2\pi\sqrt{E/(3\ln E)}$. In the paper by 
Vaughan \cite{vaughan} the prefactor has also been given by calculating 
$\sqrt{2\pi S^{(2)}(\beta)}$ which agrees with the calculation given here.

Next we consider corrections to the the asymptotic result given in Eq.\ (\ref{partlo}).

\subsection{Higher-order corrections}

The results of the previous subsection may be further improved by including additional terms 
that were neglected so far. This is done by assuming the saddle-point solution to be of the 
form 
\beq
\beta_0
=\pi\sqrt{\frac{1}{3E\ln(E)}}
\left[1+a\,\frac{\ln[\ln(E)]}{\ln(E)}+b\,\frac{1}{\ln(E)}\cdots\right],
\eeq
where $a,b$ are arbitrary coefficients to be determined using the equation above up to order 
$1/\!\ln(E)$. The form 
of the solution is suggested by the transcendental equation (\ref{trans}) itself. Since the 
LHS of (\ref{trans}) is a monomial in $E$, the only way this can be satisfied is to have 
additional corrections to cancel the non-leading terms. The saddle-point condition to go beyond 
the leading order is given in Eq.\ (\ref{spcond}) which is rewritten more conveniently as
\beq
E=-\frac{f_1}{\beta_0^2\ln\beta_0}\left[1+\frac{1-f_2/f_1}{\ln\beta_0}
+O(1/\ln^2\beta_0)\right].
\label{spcondnlo}
\eeq
We expand the unknowns on the RHS to the desired order $1/\!\ln(E)$ in the limit of large $E$.
$$\beta_0^2=\frac{\pi^2}{3E\ln(E)}
\left[1+2a\,\frac{\ln[\ln(E)]}{\ln(E)}+2b\,\frac{1}{\ln(E)}+O\{1/\!\ln^2(E)\}\right],$$
$$
\ln(\beta_0)=-\frac{1}{2}\ln(E)\left[1+\frac{\ln[\ln(E)]}{\ln(E)}
-\ln\left(\frac{\pi^2}{3}\right)\frac{1}{\ln(E)}
-2a\,\frac{\ln[\ln(E)]}{\ln^2(E)}-2b\,\frac{1}{\ln^2(E)}\right].
$$
and 
$$\beta_0^2\ln(\beta_0)=-\frac{\pi^2}{6E}
\left[1+(2a+1)\frac{\ln[\ln(E)]}{\ln(E)}+\frac{2b-\ln(\pi^2\!/3)}{\ln(E)}
+O\{1/\!\ln^2(E)\}\right]\,,
$$
Substituting these in Eq.\ (\ref{spcondnlo}) we determine the coefficients
$a,b$ as 
$$a=-\frac12\,, \quad b=\ln(\pi/\sqrt{3})+\left(\frac{f_2}{f_1}-1\right),$$ 
and therefore
\beq
\beta_0
=\pi\sqrt{\frac{1}{3E\ln(E)}}
\left[1-\frac{1}{2}\,\frac{\ln[\ln(E)]}{\ln(E)}+
\frac{\ln(\pi/\sqrt{3})+(f_2/f_1 -1)}{\ln(E)}\cdots\right].
\label{solnnlo}
\eeq

The density of prime partitions is then obtained by substituting the above 
solution into
\beq
\rho(E) = \frac{\exp\left[2\beta_0 E\left(1+\frac{(\beta_0 E)^2}{2f_1 E}\right)+\cdots\right]} 
          {\sqrt{2\pi(2\beta_0 E/\beta_0^2)
          \left[1-\frac{(\beta_0 E)^2}{2f_1E}+\cdots\right]}}\,, 
\eeq
where we have kept the NLO term in the density consistent with the order to which the
solution has been obtained.  
Substituting $\beta_0 E$ from Eq. (\ref{solnnlo}) we finally obtain
\beq
\rho(E) = \frac{\exp \left\{ {2\pi\sqrt{\frac{E}{3\ln(E)}}}
\left[1-\frac{1}{2}\frac{\ln[\ln(E)]}{\ln(E)}+
\frac{[f_2/f_1+\ln(\pi/\sqrt{3})]}{\ln(E)}+\cdots\right]
\right\}} {\sqrt{\{4[3\ln(E)]^{1/2}E^{3/2}+\cdots\}}}\,.
\label{partnlo}
\eeq
Identifying $\rho(E)$ with ${\cal P}(n$=$E)$, the above equation gives the asymptotic prime
partitions of an integer $n$. The first correction to the exponent given above, proportional to
$\ln[\ln(E)]/\!\ln(E)$, is similar to that given by Vaughan \cite{vaughan} except that its coefficient 
here is $-\frac12$ instead of +1. In the following section we shall test the approximation
obtained by ignoring all higher-order terms indicated by the dots above, thus defining
\beq
{\cal P}_{as}(n) = \frac{1}{2\,[3\ln(n)]^{\frac14}\,n^{\frac34}}\,
                   \exp \left\{ 2\pi\sqrt{\frac{n}{3\ln(n)}} 
                   \left[1-\frac{1}{2}\frac{\ln[\ln(n)]}{\ln(n)}
                   +\frac{[f_2/f_1+\ln(\pi/\sqrt{3})]}{\ln(n)}\right]\!\right\}.  
\label{partas}
\eeq
This is the main result of our paper. A few comments might be in order here, before
we compare our result with exact numerical values.

\begin{itemize}

\item The leading term in the exponent, namely $2\pi\sqrt{\frac{n}{3\ln(n)}}$, agrees with 
the previously known results \cite{roth,yang,vaughan}.

\item The prefactor given by $\{2\,[3\ln(n)]^{\frac14}\,n^{\frac34}\}^{-1}$ agrees also
with that given by Vaughan \cite{vaughan}.

\item The first correction term to the exponential, given by 
$-\frac{1}{2}\frac{\ln[\ln(n)]}{\ln(n)}$, has also been calculated
by Vaughan \cite{vaughan} but with a coefficient +1, instead of our coefficient $-\frac12$
which we believe is its correct value.

\item While Vaughan \cite{vaughan} has only given an estimate of the remaining error beyond 
the first correction in the exponential, we have been able to determine the exact coefficient
of the successive term $\propto 1/\!\ln(n)$ in the exponential. As far as we know, 
this term has not been given in the literature so far.
\end{itemize}

\newpage

\section{Numerical studies of the asymptotic function ${\cal P}_{as}(n)$}

\medskip

\subsection{Evaluation of exact data base for ${\cal P}(n)$}

We evaluate the prime partition ${\cal P}(n)$ using a standard method. Given an 
integer $n$, find the distinct primes that divide $n$. The sum of distinct 
prime factors that decompose n is denoted by $\mathscr{S}(n)$ \cite{oeis}. 
For example, $\mathscr{S}(4) = 2$ since $4 = 2 \cdot 2$ has only one distinct 
prime that divides it; $\mathscr{S}(6) = 5$ since $6 = 2 \cdot 3$, or 
$\mathscr{S}(52) = 15$ since $52 = 2 \cdot 2 \cdot 13$. (Note: if a prime 
factor occurs several times, it should only be counted once.) Once the sum of 
prime factors $\mathscr{S}(n)$ is generated in a table, the following 
recursion relation \cite{SP95} is used to compute the prime partitions 
(without any restriction)
\beq
  {\cal P}(n) = \frac{1}{n} \left[\mathscr{S}(n) + \sum_{k=1}^{n-1} \mathscr{S}(k) 
          \cdot {\cal P}(n-k)\right],
\label{pexact}\eeq
which involves all prime partitions of integers less than $n$. This procedure 
is very time consuming for large $n$. We have been able to compute ${\cal P}(n)$ for 
$n$ up to 8,654,775. But, as we shall see, even this large number is not 
sufficient to reach the asymptotics of ${\cal P}(n)$.

\subsection{Numerical study of ${\cal P}_{as}(n)$}

Using the above derived data base for the exact ${\cal P}(n)$, we now test various
approximations for their asymptotic behavior. Rather than calculating the exponentially
growing full function ${\cal P}(n)$, we look at its logarithm. We compare numerically the 
logarithm of the exact ${\cal P}(n)$ with that of the following approximations:

\begin{itemize}

\item To lowest order (LO), we set the prefactor of the exponent in (\ref{partas}) to 
unity, ignoring its denominator, and just keep the leading exponential term
\beq
{\cal P}_0(n) = \exp\left\{2\pi\sqrt{\frac{n}{3\ln(n)}}\right\},
\label{lnp0}
\eeq
an asymptotic result that has been known for a long time \cite{roth,yang}.

\item The next expression is that of Vaughan \cite{vaughan}:
\beq
{\cal P}_V(n) = \frac{1}{2[3\ln(n)]^{1/4}n^{3/4}}\,
                \exp \left\{ 2\pi\sqrt{\frac{n}{3\ln(n)}} 
                \left[1+\frac{\ln[\ln(n)]}{\ln(n)}\right] \right\}.  
\label{lnpv}
\eeq
We repeat the fact that the NLO correction term in the exponent here has a different coefficient (+1) 
from the coefficient ($-\frac12$) in our result (\ref{partas}).

\item The third approximation we investigate is our asymptotic result (\ref{partas}) 
derived in the previous section.

\end{itemize}

The numerical comparison of the above three expressions with the exact prime partitions is now 
discussed in several steps.

We first plot $\ln {\cal P}(n)$ versus $n$ for the various approximations in Fig.\ \ref{figp1}.
The solid (black) curve gives the exact values $\ln {\cal P}(n)$. Our present approximation 
(\ref{partas}), shown by the dashed (red) line, comes closest to it, improving noticeably 
over the LO approximation $\ln {\cal P}_0(n)$ (\ref{lnp0}), shown by the dash-dotted 
(blue) line, in that the remaining error is reduced by about a factor of two for $n\simg 10^6$. 
The expression (\ref{lnpv}) of Vaughan, shown by the dotted (green) curve, overshoots the 
exact values substiantially and is actually much worse than the LO approximation -- which does 
not appear to have been noticed so far. 

\begin{figure}[h]
\centering
\vspace*{-1.3cm}
\includegraphics[angle=0,width=0.8\columnwidth,clip=true]{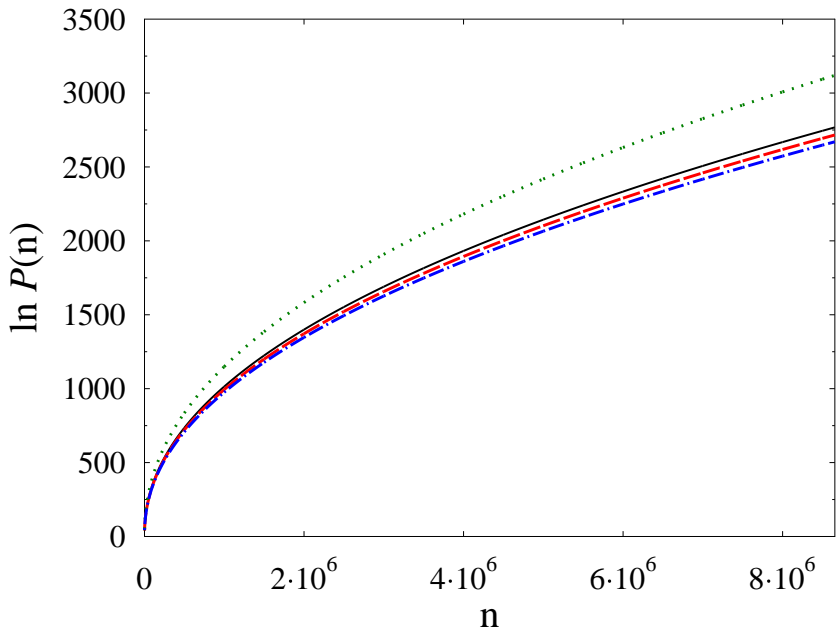}
\vspace*{-.9cm}
\caption{Logarithms $\ln {\cal P}(n)$ in various approximations. Solid line (black): exact numerical values.
Dashed (red): $\ln {\cal P}_{as}(n)$ (\ref{partas}), dash-dotted (blue): LO $\ln {\cal P}_0(n)$ (\ref{lnp0}), 
dotted(green): Vaughan (\ref{lnpv}).} 
\label{figp1} 
\end{figure} 

\begin{figure}[h]
\centering
\vspace*{-1.0cm}
\includegraphics[angle=0,width=1\columnwidth,clip=true]{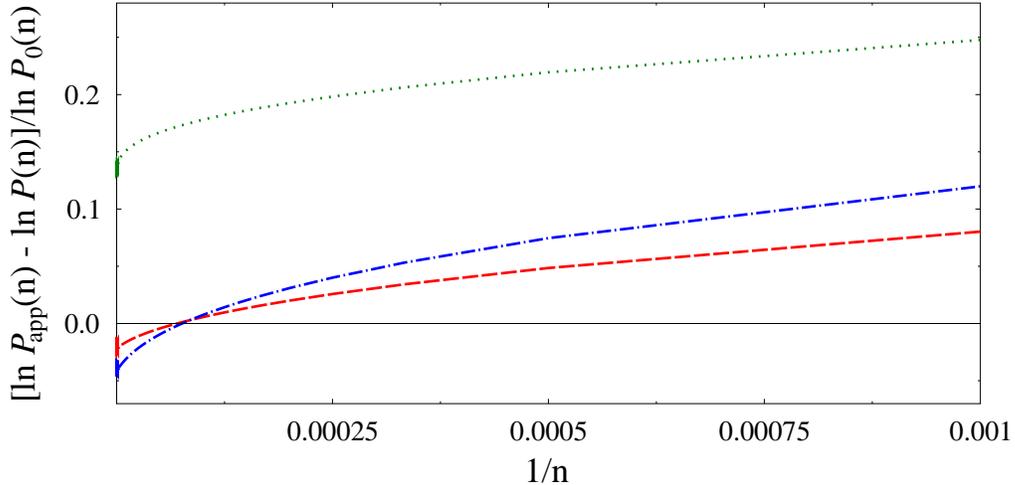}
\vspace*{-1.2cm}
\caption{Relative differences $[\ln {\cal P}_{app}(n)-\ln {\cal P}(n)]/\ln {\cal P}_0(n)$ plotted 
versus $1/n$. Dashed (red): present (\ref{partas}), dash-dotted (blue): LO term (\ref{lnp0}),
dotted (green): Vaughan (\ref{lnpv}).} 
\label{figp3} 
\end{figure} 

From this figure we can, however, not assess how the various approximations approach the correct 
asymptotics, since all curves increase monotonously. 
To this purpose we next show in Fig.\ \ref{figp3} the relative differences of 
the approximated logarithms, $[\ln {\cal P}_{app}(n)-\ln {\cal P}(n)]/\ln {\cal P}_0(n)$, 
and plot them versus $1/n$ so that they should tend to zero for $n\to \infty$ (i.e., towards 
the left vertical axis in the figure). Shown are, with the same symbols (and colors) as
above, our present approximation (\ref{partas}), the leading term (\ref{lnp0}), and that 
of Vaughan (\ref{lnpv}), all in the region $0 \leq 1/n \leq 0.001$ (i.e., $n>1000$). 
Here we see that sign changes occur in the two lowest curves: at $n\sim$ 5,800 for (\ref{partas}), 
and at $n\sim$ 13,000 for (\ref{lnp0}). They hence approach zero from below, 
while the curve of Vaughan (\ref{lnpv}) stays far up on the positive side.

\begin{figure}[h]
\centering
\vspace*{-0.9cm}
\includegraphics[angle=0,width=0.95\columnwidth,clip=true]{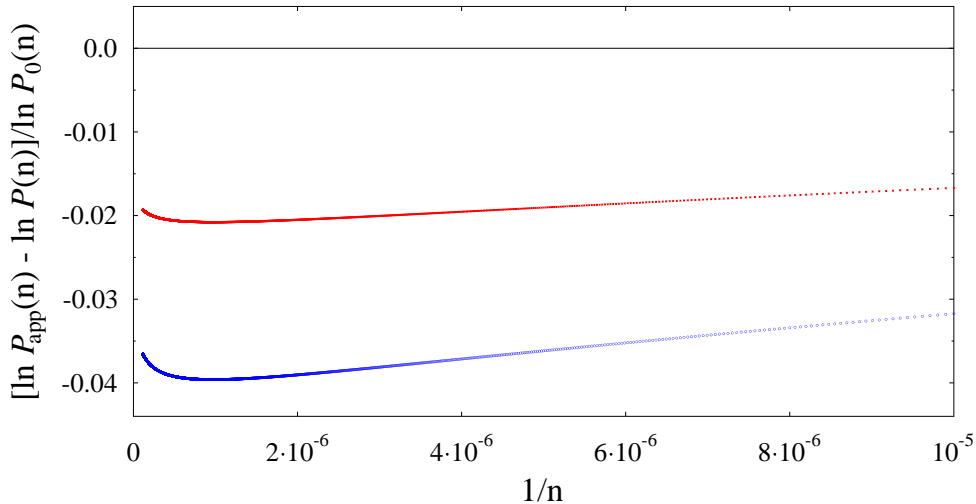}
\vspace*{-0.6cm}
\caption{Same as Fig.\ \ref{figp3} in the lowest region $1/n \leq 10^{-5}$. Upper curve
(red): our result (\ref{partas}); lower curve (blue): LO result (\ref{lnp0}). Vaughan 
curve not seen at this scale.} 
\label{figp4} 
\end{figure} 
\vspace*{-0.2cm}

In order to see to which extent the two lower curves approach the asymptotic result 0, we now focus 
on the largest values of $n$ available in our computation and further reduce the scale to $1/n \leq 
10^{-5}$, shown in Fig.\ \ref{figp4}. Vaughan's curve cannot be seen at this scale. Clearly, our 
result reduces the remaining error of the LO approximation by nearly a factor two in this region. We 
also notice that around $n\sim 10^6$, the curves have reached a minimum and then bend upwards, so 
that they do appear to go asymptotically to zero. However, the differences are still finite even 
for our largest value $n_{max}=$ 8,654,775. We must therefore question how far one has to go to reach
the correct asympototics ${\cal P}(n)$. Although the two curves in Fig.\ \ref{figp4} clearly bend 
up towards zero for $1/n\to 0$, the slopes at $n_{max}$ are such that there may be still be a long 
way to go -- too long perhaps to be covered by any numerical computation of the exact ${\cal P}(n)$.

We conclude that our result ${\cal P}_{as}(n)$ in (\ref{partas}) does appear to have the correct 
asymptotic behaviour, but that even the included corrections beyond the LO are not sufficient to 
reach the exact partitions in our numerically accessible region.


\section{Summary and conclusions}

In this paper we have discussed ${\cal P}(n)$, the number of ways a given integer may be written 
as a sum of primes -- a central theme in number theory. We have adopted methods used in quantum 
statistical mechanics where the central problem is the number of ways in which energy is distributed 
among particles occupying single-particle states. The partition function in statistical 
mechanics plays the role of the generating function of partitions. We have applied this method
to the problem of prime partitions of an integer. The dominant integral is evaluated using the 
saddle-point method.

\medskip

The main results of the paper may be summarised as follows:

\begin{itemize}

\item While the leading-order (LO) asymptotic form Eq.\ (\ref{lnp0}) has been known 
for some time, we derive non-leading order (NLO) corrections to the exponent. There 
has not been much discussion in the literature on the prefactor to the exponential form 
(\ref{lnp0}), nor of the NLO corrections. An exception is Vaughan \cite{vaughan} who 
derived the same prefactor and also a NLO correction to the exponent in (\ref{lnp0}). We 
obtain a NLO contribution to the exponent of the same form, but with a different coefficient 
($-\frac12$) than that of Vaughan \cite{vaughan} (+1). Our coefficient brings a considerable 
improvement of the asymptotics compared to that of Vaughan. 

\item We also obtain a higher order correction beyond NLO which, to the best of our knowledge, 
is not available in the literature. 

\item We use a well-known algorithm to compute the exact prime partitions, in order to 
compare analytical expressions for asymptotic prime partitions numerically. We have been 
able to do this up to more than 8 million in $n$. To our knowledge, a numerical comparison 
of exact results with asymptotic expressions has not been done up to this range before now. 
This is presumably also the reason why it has not been noticed so far that Vaughan's 
asymptotic expression (\ref{lnpv}) actually performs far worse than the lowest-order (LO) 
approximation (\ref{lnp0}) for $n \simg 10^6$ (see Figs.\ \ref{figp3},\ref{figp4}).

\item It has been known from earlier work (see for example Ref.\ \cite{muoi}) that for 
partitions of integers into integers, the asymptotic expressions for $p(n)$ are reached
very rapidly -- for $n$ of the order of $100$ or more. This 
is so because, as shown by Radmacher \cite{rad}, the exact expression for integer partitions 
may be written as a convergent series. The $k$-th term in the series is of order 
$\exp\left(\frac{\pi}{k}\sqrt{\frac{2n}{3}}\right)$. The leading term with $k=1$ gives the 
Hardy-Ramanujan result. The first correction to the exponent is at $k=2$ and therefore 
\beq
p(n)\approx C_1(n)\, \exp\left(\pi \sqrt{\frac{2n}{3}}\right)\left[1+ C_2(n)\,
\exp\left({-\frac{\pi}{2}\sqrt{\frac{2n}{3}}}\right)\right],
\eeq
where $C_1$ and $C_2$ are $n$-dependent prefactors.
The correction to the exponent falls off exponentially. However as seen from Eq.\ (\ref{partas}),
the correction in the case of primes falls off logarithmically which explains why the asymptotic 
limit is reached much more slowly for prime partitions, as compared to that of $p(n)$.
 
\item Although both the exact ${\cal P}(n)$ and the asymptotic form ${\cal P}_{as}(n)$ given in  
(\ref{partas}) are monotonously increasing, their difference is not monotonic. In fact, we 
show that ${\cal P}_{as}(n)$ crosses ${\cal P}(n)$ around $n\sim$ 5,800 and approaches it
from below for $n\to\infty$ (within the limits of our data). The remaining error has a maximum
absolute value around $n\sim 10^6$ (see Fig.\ \ref{figp4}), beyond which it clearly tends towards zero.

\item Our main conclusion is that our result ${\cal P}_{as}(n)$ given in (\ref{partas}) improves clearly
over the the LO expression ${\cal P}_0(n)$ in (\ref{lnp0}), it appears to have the correct asymptotic 
behaviour for $n\to \infty$, but that even the corrections beyond the LO are not sufficient 
to reach the exact ${\cal P}(n)$ in the numerically accessible region.

\end{itemize}

Concerning partitions of integers $n$ into smaller integers or into squares of integers, there exist
physical quantum Hamilonians which lead to these partitions, namely the harmonic oscillator and the
infinite square-well potential \cite{muoi}. Regarding the spectrum of primes, there have been attempts to
construct potentials whose eigenvalues are the primes. Unfortunately, these potentials keep changing
upon inclusion of more primes and have a fractal-like character \cite{muss,seka,zyl}. That the prime
spectrum can be reproduced from the non-trivial zeros of the Riemann zeta function is shown in the
trace formula given in Eq.\ (\ref{gxsc}) of the Appendix and illustrated in Fig.\ \ref{trf}. As we
have mentioned already in the beginning of Sec.\ II.A, semiclassical trace formulae can give insights 
into deep-lying mathematical connections. On the physical side they provide the connection between a 
quantum spectrum and the periodic orbits of the corresponding classical Hamiltonian, and on the purely 
mathematical side, they connect spectral theory with symplectic geometry (in particular, with 
geodesics on Lagrangian manifolds) \cite{gutz}.

\bigskip
\bigskip

{\bf \small{Acknowledgements}}


The authors would like to thank Shouvik Sur (Florida State University, USA), 
Rajesh Ravindran (The Institute of Mathematical Sciences, Chennai, India), 
and Ken-ichiro Arita (Nagoya Insitute of Technology, Japan) for collaboration 
and stimulating discussions in the initial stages of this work. JB, MB and RKB thank 
the Institute of Mathematical Sciences where part of this work was done, for 
its hospitality. MVN thanks the Department of Physics and Astronomy, McMaster 
University, for its hospitality during the final stages of this work. Finally, we
want to express our gratitude to Florin Spinu who, in a private communication,
has not only confirmed our coefficient $-\frac12$ of the next-to-leading term in the exponent
of ${\cal P}_{as}(n)$, but also located the error in Ref.\ \cite{vaughan} which led to
the wrong coefficient +1 of that term.

\appendix*

\section{Some details about the density of primes} 

In this section, we discuss two approximations to the density of primes $g(x)$ defined in
(\ref{gofx}), which is related to the function $\pi(x)$ that counts the number 
of primes $p \leq x$ by a differentiation:
\beq
g(x) = \frac{d\pi(x)}{dx}\,.
\label{gpip}
\eeq
Both $\pi(x)$ and $g(x)$ have been the object of a lot of research in number theory. $\pi(x)$ is
a stair-case function whose average part is given by the asymptotic form
\beq
\pi(x)  \sim   \frac{x}{\ln(x)}\,,
\eeq
which is a consequence of the prime number theorem. A more refined asymptotic form is (see, e.g., \cite{edw}):
\beq
\pi(x) \sim \frac{x}{\ln(x)} + \frac{x}{[\ln(x)]^2} + \dots + (n-1)!\,\frac{x}{[\ln(x)]^n}\,.
\label{pias}
\eeq
Differentiating it yields the asymptotic expression for the density of primes
\beq
g(x)  \sim  1/\ln(x)\,,
\label{gofxas}
\eeq
whereby all higher-order terms coming from (\ref{pias}) have cancelled successively. In Sec.\ III
we have used the above asymptotic form for the average prime density $g_{av}(x)$.
 
An alternative expression for $\pi(x)$ may be derived from a function studied by Riemann in 1859,
called $J(x)$ and further discussed by Edwards \cite{edw}
\beq
J(x) = \sum_{n=1}^\infty \frac{1}{n}\,\sum_{p} \,\Theta(x-p^n) \,.  \qquad (x>0)
\label{Jx}
\eeq
Here $p$ runs over all primes and $n$ over all integers, and $\Theta(x)$ is the standard step 
function: $\Theta(x)=1$ for $x\geq 0$ and $\Theta(x)=0$ for $x<0$. In a seminal paper \cite{riemann},
Riemann gave an expression for $J(x)$ in terms of zeros of the zeta function. We use his 
expression and employ the Moebius inversion formula (cf. \cite{edw})
\beq
\pi(x)=\sum_{m=1}^\infty \frac{\mu(m)}{m}\,J(x^{1/m})\,,  
\label{pip}
\eeq
where $\mu(m)$ is the Moebius function [$\mu(1)=1$]. Taking the derivative according to (\ref{gpip}),
we obtain the following expression for the density of primes (given also in \cite{berry})
\beq
g_{sc}(x) = \frac{1}{x\ln x} \sum_{m=1}^\infty \frac{\mu(m)}{m}
              \left[\,x^{1/m}-\frac{1}{(x^{2/m}-1)}
              -2\,x^{1/2m}\sum_\alpha \cos\left(\frac{\alpha}{m}\ln x\right)\right].
\label{gxsc}
\eeq
Here $\alpha>0$ are the non-trivial zeros of the Riemann zeta function along the positive half-line, 
and the validity of the Riemann hypothesis has been assumed. This expression, which does not appear 
to be widely known, has the form of a semiclassical ``trace formula'' \cite{book,gutz} and we have 
therefore denoted it with the subscript ``$sc$'' for ``semiclassical''. Ideally, $g_{sc}(x)$ should 
yield the exact prime density $g(x)$ in (\ref{gofx}) if the sum over $\alpha$ is not truncated, and if
the Riemann hypothesis is true. 

We have tested Eq.\ (\ref{gxsc}) numerically in order to convince
ourselves of its validity. For practical purposes, we have coarse-grained it, replacing
the delta functions in (\ref{gofx}) by normalized Gaussians with a width $\gamma$, and
correspondingly coarse-grained Eq.\ (\ref{gxsc}) as described in Sec.\ 5.5 of \cite{book}. 
Fig.\ \ref{trf} shows the results, obtained using the lowest 3000 Riemann zeros $\alpha$.
We see that the coarse-grained trace formula indeed reproduces the Gaussian-smoothed density 
of primes, replacing the delta functions in (\ref{gofx}) by Gaussians centered exactly at 
the primes $p$. (Note that the sum over $m$ can be truncated for any finite value of $x$; in 
the situation described here, $m_{max}=14$ was sufficient.)

We note that the average part of $g_{sc}(x)$ in (\ref{gxsc}) is not suitable for use in Eq.\
(\ref{zav}), because it has a pole structure that cannot be integrated easily. Numerically
we found it to be very well approximated by the familiar asymptotic expression (\ref{gofxas}).

\begin{figure}
\centering
\vspace*{-2.1cm}
\hspace*{-1.6cm}\includegraphics[width=1.25\columnwidth,clip=true,angle=0]{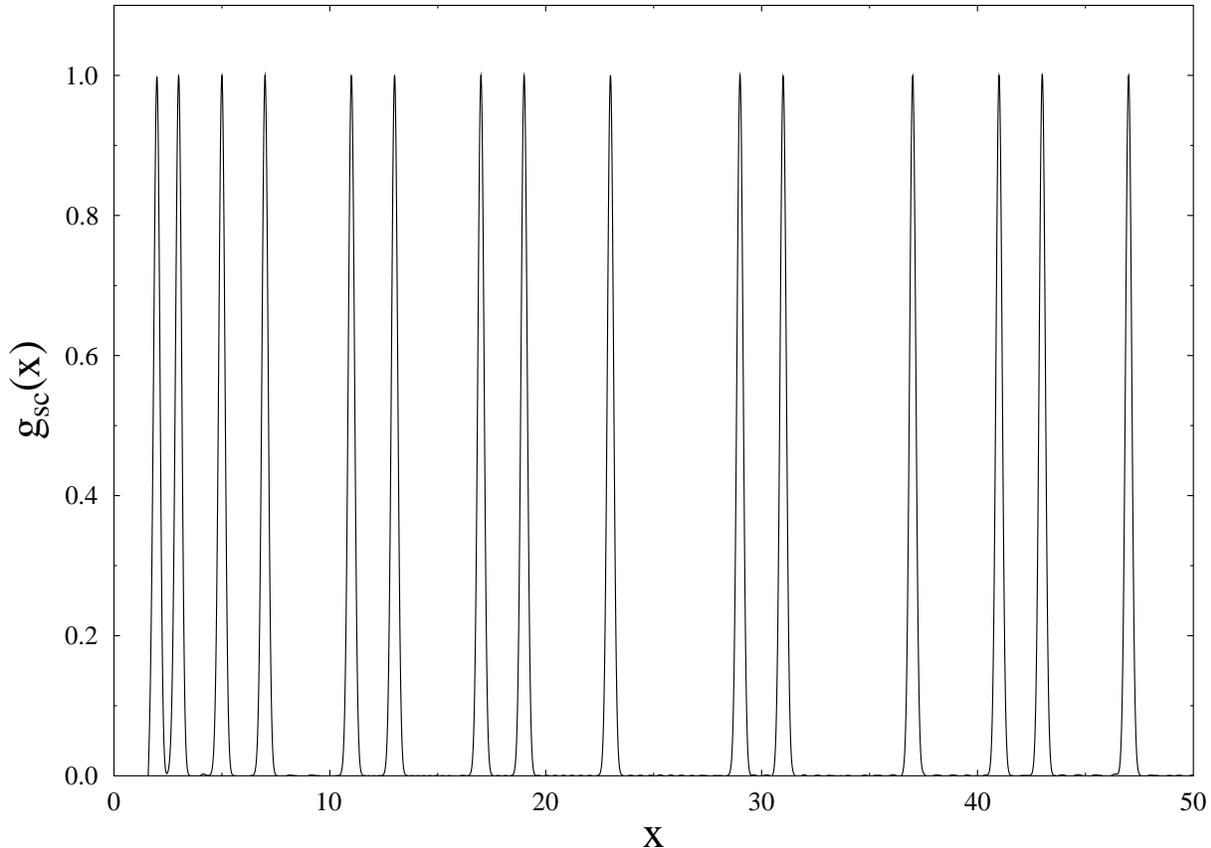}
\vspace*{-2.2cm}
\caption{Density of primes $g(x)$ obtained by the semiclassical expression $g_{sc}(x)$ in (\ref{gxsc}),
using the lowest 3000 Riemann zeros $\alpha$ and $m_{max}=14$, coarse-grained with a Gaussian 
width $\gamma_{sh}=0.1$.}
\label{trf} 
\end{figure} 

\end{document}